\documentclass[pra,aps,twocolumn,amsmath,amssymb,superscriptaddress]{revtex4}
\usepackage{bm,graphicx,amsmath}
\usepackage{bbm}

\usepackage{bm}
\usepackage{amssymb}
\usepackage{epsfig}
\usepackage{epstopdf}
\usepackage{amstext}
\usepackage{latexsym}
\usepackage{hyperref}
\usepackage{times}
\usepackage{color}

\newcommand{\da}{^\dagger}
\newcommand{\ket}[1]{\left\vert#1\right\rangle}

\begin{document}

\title{Characterization of Bose-Hubbard Models with Quantum Non-demolition Measurements}

\author{B. Rogers}
\affiliation{Centre for Theoretical Atomic, Molecular and Optical Physics, Queen's University Belfast, Belfast BT7 1NN, United Kingdom}
\author{M. Paternostro}
\affiliation{Centre for Theoretical Atomic, Molecular and Optical Physics, Queen's University Belfast, Belfast BT7 1NN, United Kingdom}
\author{J. F. Sherson}
\affiliation{Department of Physics and Astronomy, Ny Munkegade 120, Aarhus University, 8000 Aarhus C, Denmark}
\author{G. De Chiara}
\affiliation{Centre for Theoretical Atomic, Molecular and Optical Physics, Queen's University Belfast, Belfast BT7 1NN, United Kingdom}

\begin{abstract}
We propose a scheme for the detection of quantum phase transitions in the 1D Bose-Hubbard (BH) and 1D Extended Bose-Hubbard (EBH) models, using the non-demolition measurement technique of quantum polarization spectroscopy.  We use collective measurements of the effective total angular momentum of a particular spatial mode to characterise the Mott insulator to superfluid phase transition in the BH model, and the transition to a  density wave state in the EBH model.  We extend the application of collective measurements to the ground states at various deformations of a super-lattice potential.
\end{abstract}

\date{\today}

\maketitle

\noindent
\section{Introduction}

The ability to trap and cool atoms in optical lattices~\cite{review_optical_lattices} has opened up new avenues in atomic physics. This has, perhaps unexpectedly, moved this area closer to condensed matter physics. Optical lattices are in fact one of the most promising candidates for the realisation of a universal quantum simulator~\cite{nature_insight}. This is a programmable quantum system that can be made to evolve according to a desired theoretical model. One of the first proposals of a quantum simulator with ultracold bosonic atoms in optical lattices was the one given in Ref.~\cite{jaksch98} where the implementation of the Bose-Hubbard model (BHM) was suggested and later implemented~\cite{Greiner2002}. This model was originally a more abstract version of the corresponding fermionic Hubbard model which describes the propagation of fermionic particles in conductors. The BHM exhibits quite interesting physics with an interaction induced insulator called the Mott-insulator state and a dissipation free conducting superfluid~\cite{fisher89}. 

Since then a plethora of proposals of quantum simulators with cold atoms has been put forward that include different lattice geometries, disorder, impurities, interactions beyond nearest-neighbour, synthetic gauge fields, spin-orbit coupling, multi-orbital BHM, etc.~\cite{dutta_review}. Besides condensed matter physics, applications in high-energy physics and controlled quantum chemistry have also been put forward \cite{BuchlerPRL2005,Jaksch2002}.

Readout of ultracold atom quantum simulators has so far mostly been achieved using time-of-flight experiments \cite{Ketterle_review} and lately also using single-site resolved microscopy~\cite{atom_microscope}. The former method provides the momentum distribution and density-density correlations of the atomic sample in the trap, thus allowing one to distinguish between Mott-insulator and superfluid states. The latter gives information about  occupations in individual sites and can therefore be used to compute  density-density correlations between specific sites \cite{endres}.
Other possible measurement schemes involve optical Bragg scattering that gives information about dynamical structure factors~\cite{corco}.

These examples of detection methods are all destructive: the sample is released from the trap as in time-of-flight experiments and the system quantum state is incoherently projected onto a Fock basis and strongly heated using single-site imaging. 

\begin{figure}[t]
\includegraphics[width=0.8\linewidth]{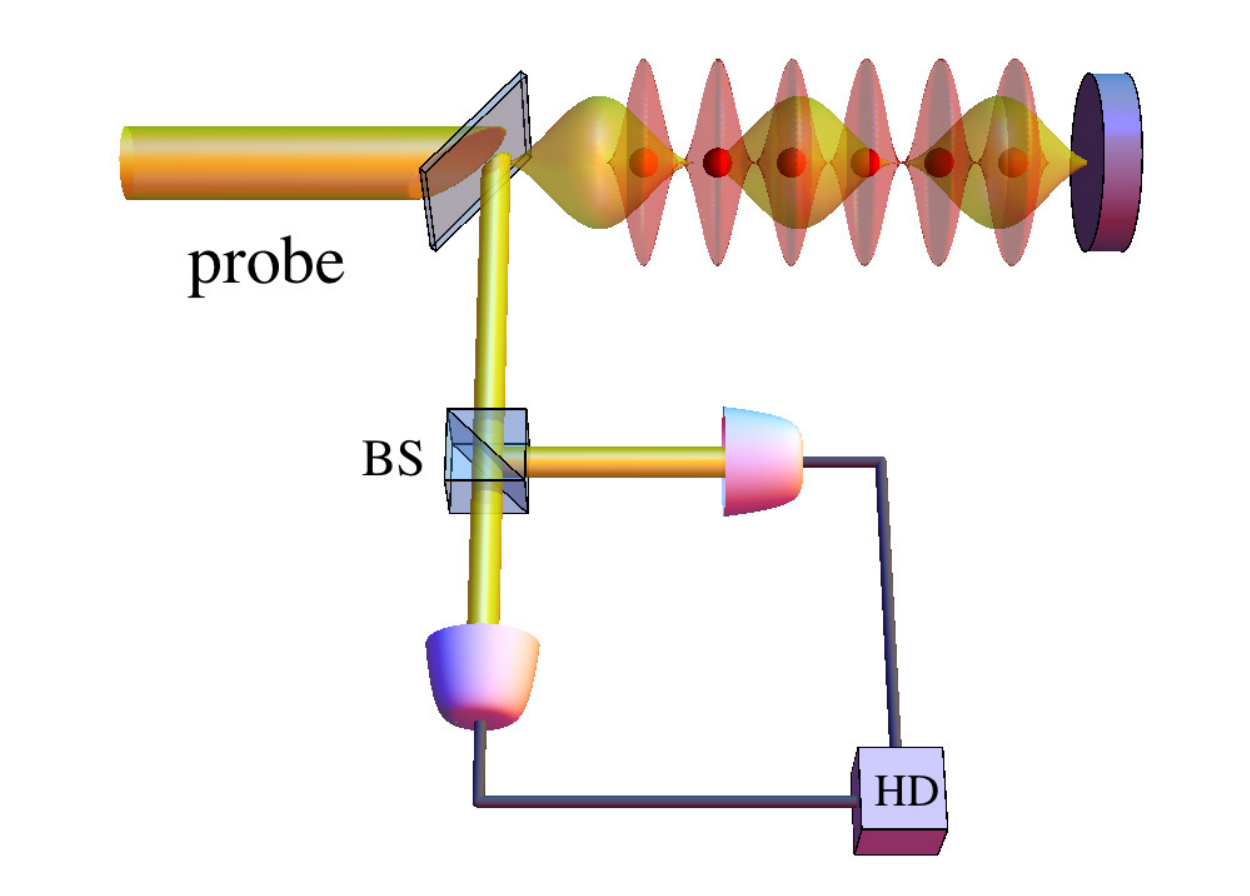}
\caption{\label{fig:setup}
(Color online) Schematic diagram of the quantum polarization spectroscopy setup: atoms are trapped in a deep optical lattice potential (red) and illuminated by a probe beam forming a standing wave over a sample of trapped atoms (yellow).  The beam is split by a beamsplitter (BS) and measured by homodyne detection (HD).
}
\end{figure}

Two alternative quantum non-demolition schemes have recently been proposed: one based on coupling lattice sites to individual cavities \cite{Ritsch}, and quantum polarization spectroscopy (QPS) \cite{eckert07,roscilde09,dechiara11}.  In this work we focus on the latter in which probing light is rotated due to the Faraday rotation induced by the magnetic moments of the illuminated atoms (see the proposed setup in Fig.~\ref{fig:setup}).

Theoretical publications have so far proven the detection of magnetic phases of spin and fermionic systems with QPS by analysing the polarization fluctuations of the light emerging from the sample~\cite{eckert07,roscilde09,dechiara11}. In particular in Ref.~\cite{dechiara11} it was shown that the output light signal can be connected to magnetic order parameters thus signalling critical points.
This detection method can even be reversed, in a sense, to enable state preparation by engineering desired spin correlations through successive operations~\cite{hauke13}.  So far, the QPS scheme has not been applied to the characterisation of quantum phase transitions in BHMs, which will be the main focus of this work.

Here we showcase the power and flexibility of this technique to characterise the ground state of the one-dimensional (1D) BHM in three distinct variants.
First, we start with the pure homogeneous 1D-BHM and we show how to employ the QPS scheme to distinguish the Mott insulator (MI) and superfluid (SF) states.
The precise determination of the Mott insulator to superfluid critical point has proven to be a hard task because commonly used techniques, such as the visibility of interference peaks in time-of-flight, give a smooth signal across the transition. Here we show, using numerical simulations, that our technique is capable of detecting not only a quantitative but also a qualitative change in behaviour across the transition allowing us to get a fairly accurate estimate of the critical point. 
Second, we consider the extended BHM (EBHM) obtained by adding a nearest-neighbour interaction term arising, for example, for dipolar atoms. In this model, two further insulating phases appear: a Haldane insulator (HI), characterised by a non-local hidden order and a density wave (DW) phase exhibiting a spatial density modulation  \cite{dallatorre,rossini12}.
 In this setting we show how to locate the critical point between the two phases. Finally, we consider bosonic atoms in a superlattice potential recently proposed for implementing atomic transistors and single and two-qubit gates~\cite{SLqubitgates}. In this setting the QPS method allows us to detect different atomic patterns arising in the superlattice.

The paper is organised as follows: In Sec.~\ref{sec:QPS} we review the fundamentals of the QPS method; in Sec.~\ref{sec:BH}, the phase transition between the MI phase and the SF phase in the BHM is found, based on the polarization of the output light emerging from the sample;  in Sec.~\ref{sec:EBH}, the phase transition between the HI phase and the DW phase in the EBHM is identified using a new measure we propose called the \textit{disparity}, which is experimentally observable.  Finally, in Sec.~\ref{sec:SL} we show how the QPS method can be used to identify sublattice arrangements in a superlattice potential and in Sec.~\ref{sec:conclusions} we draw our conclusions and briefly address the possibilities of further applications of our method.

\section{Quantum Polarization Spectroscopy} 
\label{sec:QPS}
We briefly describe the scheme of QPS by which the spatial distribution of atoms trapped in a lattice are read-out via non-demolition measurements~\cite{eckert07, roscilde09, dechiara11}, shown in Fig.~\ref{fig:setup}.  There, a probe beam propagating along the $z$ direction interacts in a standing wave configuration with the atoms trapped in an optical lattice.
An atomic collective angular momentum in the $z$-direction $\hat{J}^{\mathrm{eff}}_z$, whose form will be made explicit below, can be mapped onto the polarization state of incident light through the Faraday effect.  When far-off resonant light is incident, atomic excitations from the ground state manifold are suppressed and the excited states can be adiabatically eliminated~\cite{Kupriyanov}.  As such the effective Hamiltonian originating from the dipole interaction reads:
\begin{equation}
\hat{H}_{\mathrm{eff}}=-\kappa \hat{s}_3 \hat{J}^{\mathrm{eff}}_z,
\end{equation}
where the coupling $\kappa$ depends on the probability of resonant excitation per atom by the probe and the optical depth of the atomic sample.  The Stokes operators $\hat s_1=\frac{1}{2}(\hat a^\dagger_x \hat a_x-\hat a^\dagger_y \hat a_y),
\hat s_2=\frac{1}{2}(\hat a^\dagger_y \hat a_x+\hat a^\dagger_x \hat a_y), and 
\hat s_3=\frac{1}{2i}(\hat a^\dagger_y \hat a_x-\hat a^\dagger_x \hat a_y)$ conveniently describe the light polarization state in terms of the annihilation operator for a photon with $x$ polarization $\hat{a}_x$ or with $y$ polarization $\hat{a}_y$.  Throughout the paper we set $\hbar=1$.

We assume a strongly $x$-polarized beam, i.e. $\langle \hat{S}_1 \rangle = N_{ph}/2 \gg 1$ where $\hat S_i = \int \! \hat s_i \, \mathrm{d}t$, $i=\{1,2,3\}$ and $N_{ph}$ is the total number of photons in the beam. Using the Holstein-Primakoff transformation~\cite{HPtransformation}, we can approximate the other two Stokes operators as two effective conjugated variables: $\hat X=\hat S_2/\sqrt{N_{ph}/2}$ and $\hat P=\hat S_3/\sqrt{N_{ph}/2}$ such that 
\begin{equation}
[\hat X,\hat P] = \frac{i S_1}{N_{ph}/2} \sim i.
\end{equation}

After solving the Heisenberg equations for small times it can be found that the output quadrature of the light field, corresponding to polarization fluctuations,  can be written as
\begin{equation}
\hat{X}_{\mathrm{out}} = \hat{X}_{\mathrm{in}} - \kappa \hat{J}^{\mathrm{eff}}_z.
\end{equation}
For brevity, $\hat{J}^{\mathrm{eff}}_z$ is referred to simply as $\hat{J}$.  When the probe beam is in a standing-wave configuration (cf. Fig.~\ref{fig:setup}), properties of the atoms can be inferred with spatial resolution, defined by the spatial mode of the probing light, which is critical to the detection of quantum phases with non-trivial structures, such as the phases we discuss here.  By considering a localised single-particle basis with the use of Wannier functions, in second quantisation, the particle and site positions are discretized.  The standing-wave modulation is apparent in the expression for the effective total angular momentum
\begin{equation}\label{eq:J}
\hat{J} = \frac{2}{\sqrt{L}} \sum_{i=1}^L \cos^2\left[k(i - \alpha)\right] \hat{n}_i ,
\end{equation}
where $i$ is the lattice site index, $L$ is the lattice length, $k$ is the probe wavenumber measured in units of inverse lattice spacing, $\alpha$ is the displacement of the probe standing wave with respect to the lattice potential and $\hat{n}_i$ is the number operator of the site $i$. Here, we are assuming that the atoms are all in the same magnetic state polarized along the $z$ axis so that the effective angular momentum depends in practice on the atomic density. The mean value $\langle \hat{J} \rangle = \frac{2}{\sqrt{L}} \sum_{i=1}^L \cos^2\left[k(i - \alpha)\right]\langle \hat{n}_i \rangle$ depends on the expectation value of the density operator $\hat{n}_i$ on all sites $i$.  The variance of $\hat{J}$ is related to density-density correlations in the atomic system
\begin{align}\label{eq:varJ}
(\Delta\hat{J})^2 =~&\frac{4}{L} \sum_{i,j=1}^L \cos^2\left[k(i-\alpha)\right]\cos^2\left[k(j - \alpha)\right] \nonumber \\
&\times\left\{ \langle \hat{n}_i \hat{n}_j \rangle - \langle \hat{n}_i \rangle \langle \hat{n}_j \rangle \right\}.
\end{align}
Since the total number of particles $N$ is conserved, $(\Delta\hat{J})^2=(\Delta N)^2=0$ for $k=0$, i.e. when the whole lattice is detected by the probe beam. Throughout this work we consider unit filling with the number of atoms equal to the number of lattice sites.

For all of the following analyses we use the density matrix renormalization group (DMRG) numerical technique to find the ground states of the bosonic systems~\cite{white92,schollock05review,DMRGdechiara08}. Open-boundary conditions are used, with a maximum occupation of four bosons per site.
\begin{figure}[!ht]
	{\bf (a)}\\
	\includegraphics[width=0.9\linewidth]{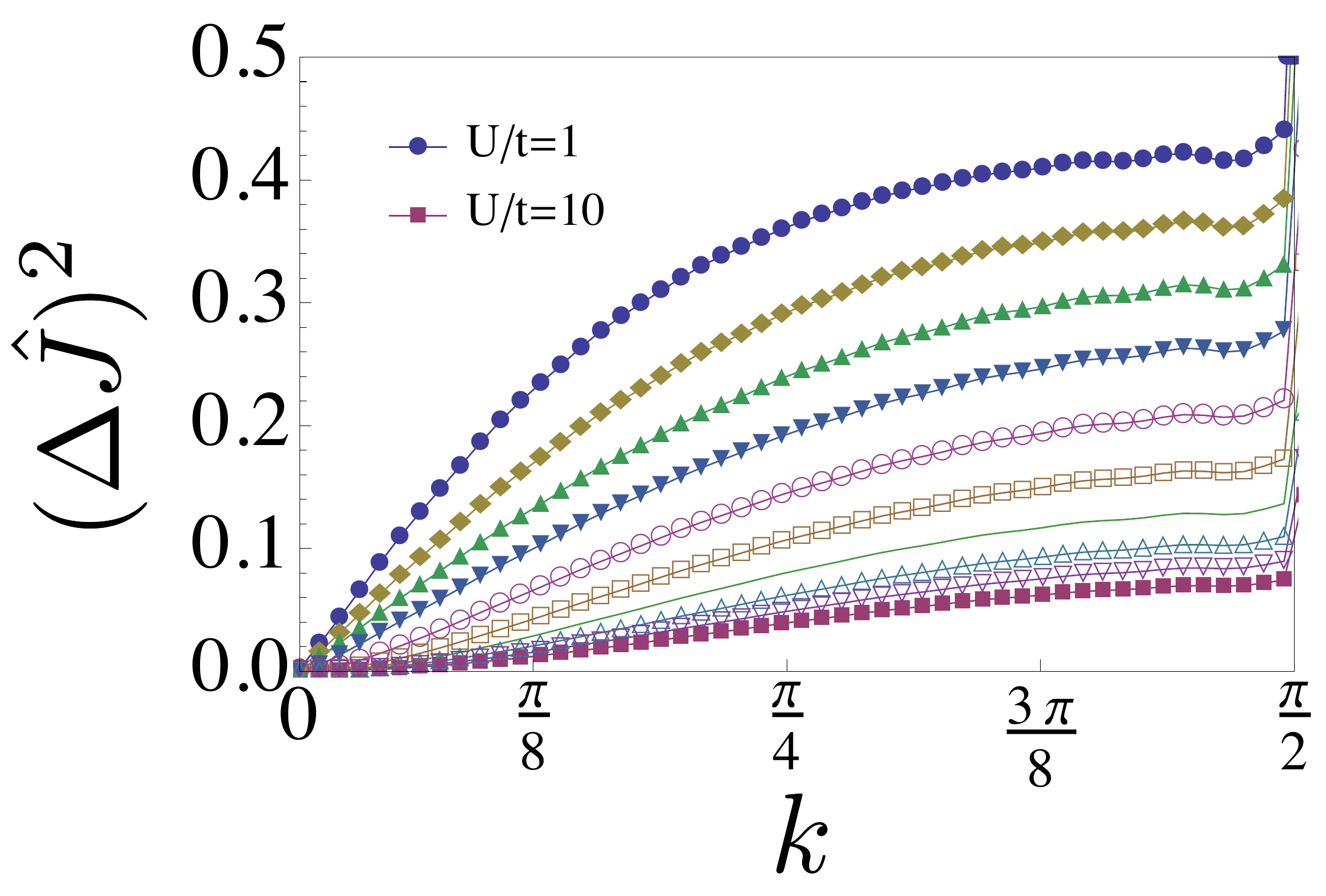}\\
	{\bf (b)}\\
\includegraphics[width=0.9\linewidth]{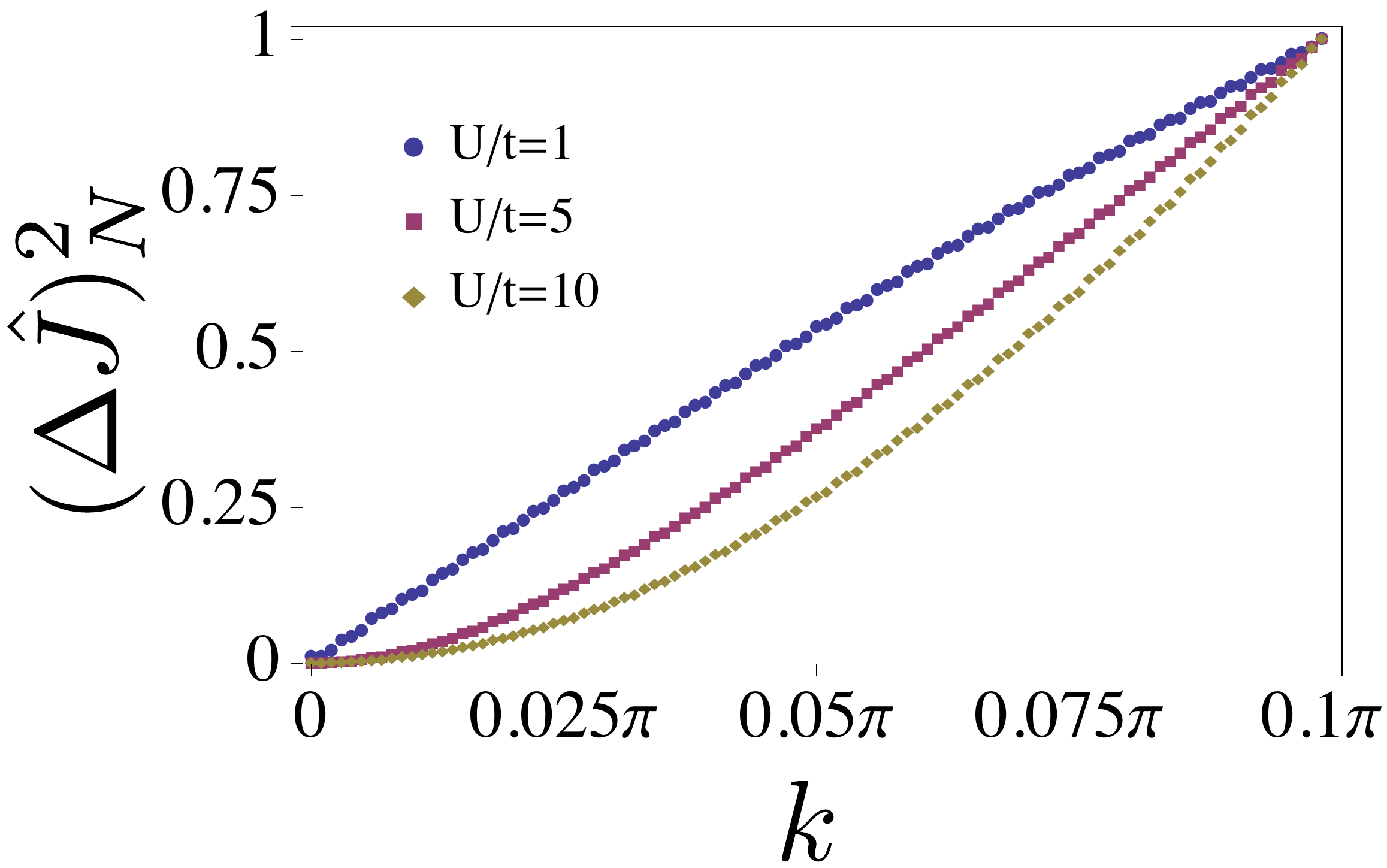}
\caption{\label{fig:JvskmultiU}
(Color online) {\bf (a)} The dependence of the variance of $\hat{J}$ on the on-site repulsion, with integer values $U/t=\{1,...,10\}$ and lattice size $L=160$. {\bf (b)} The normalised quantity $(\Delta\hat{J})^2_N$ obtained from $(\Delta\hat{J})^2$ by dividing by its value at $k=0.1\pi$  for $U/t=1,5,10$.  All quantities plotted in this figure and subsequent figures are dimensionless.
}
\end{figure}
\section{Bose-Hubbard Model: Mott insulator to superfluid transition}\label{sec:BH}
In this section, we use collective measurements of a 1D bosonic system trapped in a lattice to characterise the MI to SF quantum phase transition of the BHM.  Our DMRG calculations show that the transition can be identified by effectively comparing the variance of the lattice site populations (through $(\Delta\hat{J})^2$) in the two phases.  The Hamiltonian of the BHM is
\begin{equation}\label{eq:BHM}
\hat{H}_{BH} = -t\sum_i^L\left(\hat{b}_i^{\da}\hat{b}_{i+1} + \hat{b}_i\hat{b}_{i+1}^{\da}\right) + \frac{U}{2}\sum_i^L \left[ \hat{n}_i(\hat{n}_i -1) \right],
\end{equation}
with nearest neighbour hopping strength $t$ and on-site repulsion $U$.  The Mott insulator is the gapped phase of the BHM with integer particle filling of the lattice sites, occurring when $U$ dominates over $t$.  Such a state can be engineered by globally increasing the lattice potential. As number fluctuations in each site decrease, we expect the variance of the collective angular momentum $(\Delta\hat{J})^2$ to be suppressed. The gapless phase of the BHM is a bosonic superfluid (when $t$ dominates over $U$), with particle wavefunctions de-localised over the extent of the lattice, resulting in local number fluctuations and a corresponding increase in $(\Delta\hat{J})^2$.  
Our numerical results confirm these expectations as shown in Fig.~\ref{fig:JvskmultiU}{\bf (a)} which plots the variance $(\Delta\hat{J})^2$ as a function of the probing wave vector $k$ for $\alpha=0$ and for different values of $U/t$ both in the SF and MI phases. The critical point is at $U_c/t \approx 3.3$ \cite{kuhner00}. 

\begin{figure}[t]
\includegraphics[width=0.9\linewidth]{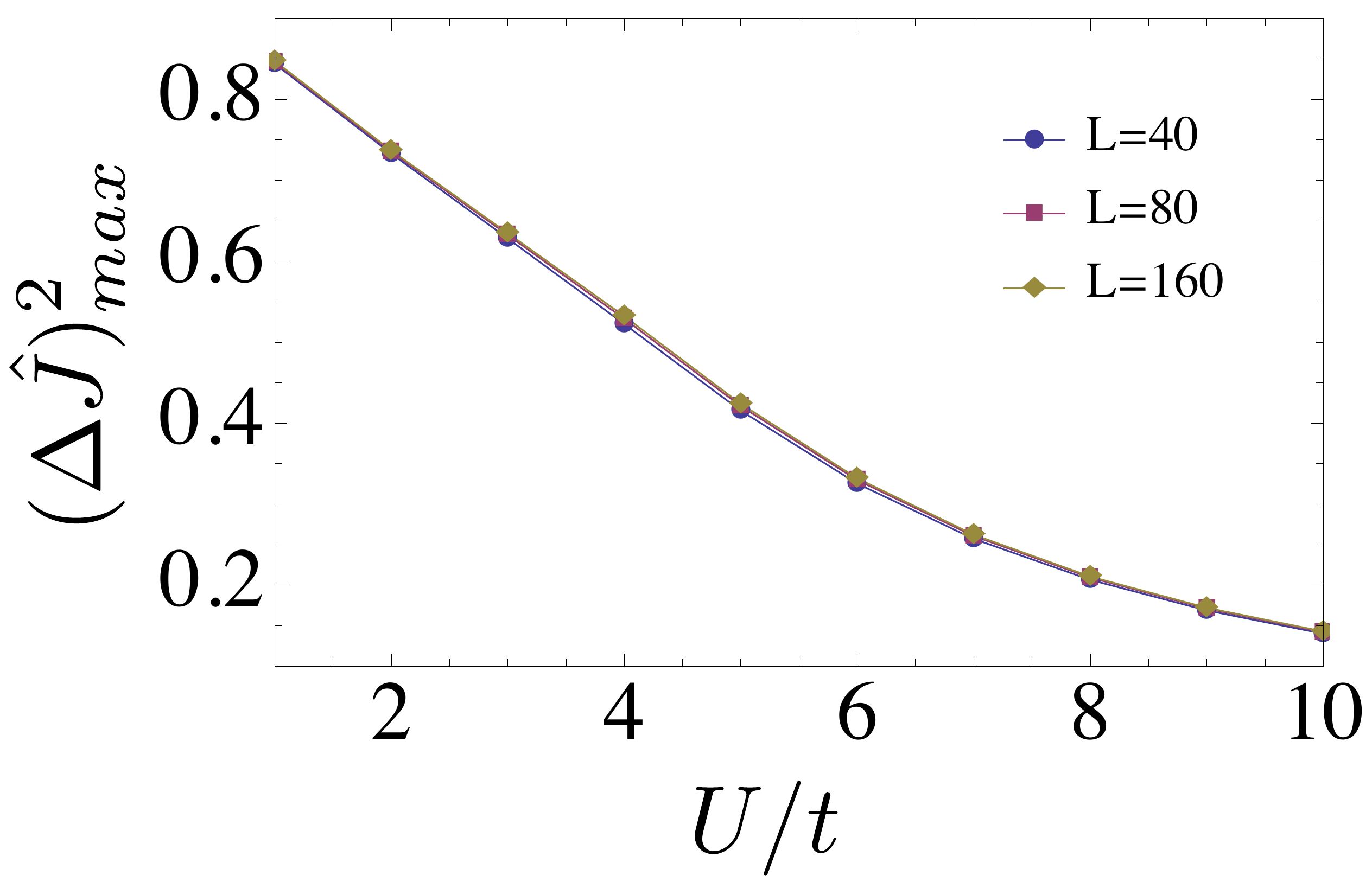}
\caption{\label{fig:JmaxvsU}
(Color online) The variance of $\hat{J}$, taken at $k=\pi/2$, against the on-site repulsion, for the lattice sizes $L={40,80,160}$.
}
\end{figure}

The peak observed in Fig.~\ref{fig:JvskmultiU}{\bf (a)} at $k=\pi/2$, corresponding to a probing lattice with twice the wavelength of the primary lattice, is a consequence of doublon-holon correlations in the ground state. Fixing $k$ to this value, at which we obtain the largest signal $(\Delta\hat{J})^2_{max}$, we observe a monotonic decrease as the system passes from a SF to a MI phase (Fig.~\ref{fig:JmaxvsU}). The quantity plotted does not depend strongly on the size of the system and can therefore be used in an experiment for benchmarking and calibrating the numerical calculations.  However, the lack of scaling of this quantity for $k=\pi/2$ prevents us from independently determining the critical point for the MI-SF transition.

To this aim we then turn to the behaviour of $(\Delta\hat{J})^2$ for small $k$. According to Luttinger liquid arguments~\cite{giamarchisbook,ejima,kuhner00} our quantity $(\Delta\hat{J})^2$, which resembles the density structure factor,   should follow a linear dependence on $k$ in the SF critical phase and a quadratic dependence in the MI phase.  

\begin{figure}[b]
\includegraphics[width=0.9\linewidth]{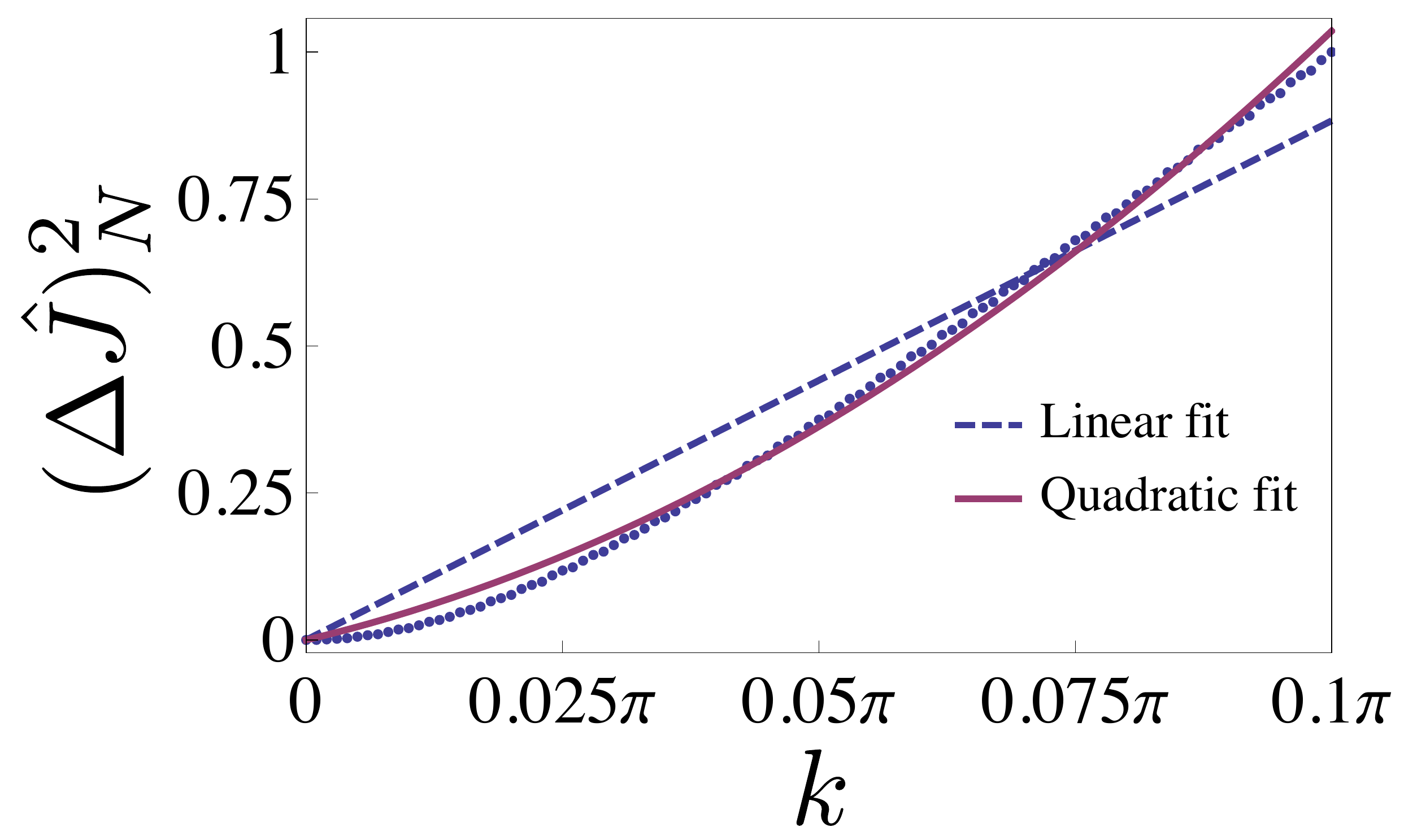}
\caption{\label{fig:U5smallk}
(Color online) For illustrative purposes, the tail of the rescaled $(\Delta\hat{J})^2$ is shown for $U/t=5$ (MI phase) and $L=160$, with  100 data points.}
\end{figure}
This dependence is indeed observed in our calculations as shown in Fig.~\ref{fig:JvskmultiU}{\bf (b)}, for  values of the on-site repulsion $1 \leq U/t \leq 10$.  We set $\alpha=0$ and focus on the range $0 \le k \le 0.1\pi$, where this distinction is more evident.  Since the magnitude of $(\Delta\hat{J})^2$ is larger in the SF phase we plot $(\Delta\hat{J})^2_N$ obtained from $(\Delta\hat{J})^2$ by dividing it by its value at $k=0.1\pi$.  To distinguish the MI and SF phases, and thereby locating the critical point, we compare the quality of a linear or quadratic fit of  $(\Delta\hat{J})^2_N$ against $k$ with zero intercept. We separately fit a linear function $y_1=a_1 k$ and a quadratic function $y_2=a_2k+bk^2$ to the variance $(\Delta\hat{J})^2_N$. One example for $U=5t$ (MI phase) is shown in Fig.~\ref{fig:U5smallk}. The plot clearly shows that the quadratic curve is a better fit than the linear ansatz. We thus repeat the same fitting for different values of $U/t$.  

\begin{figure}[t]
\includegraphics[width=0.9\linewidth]{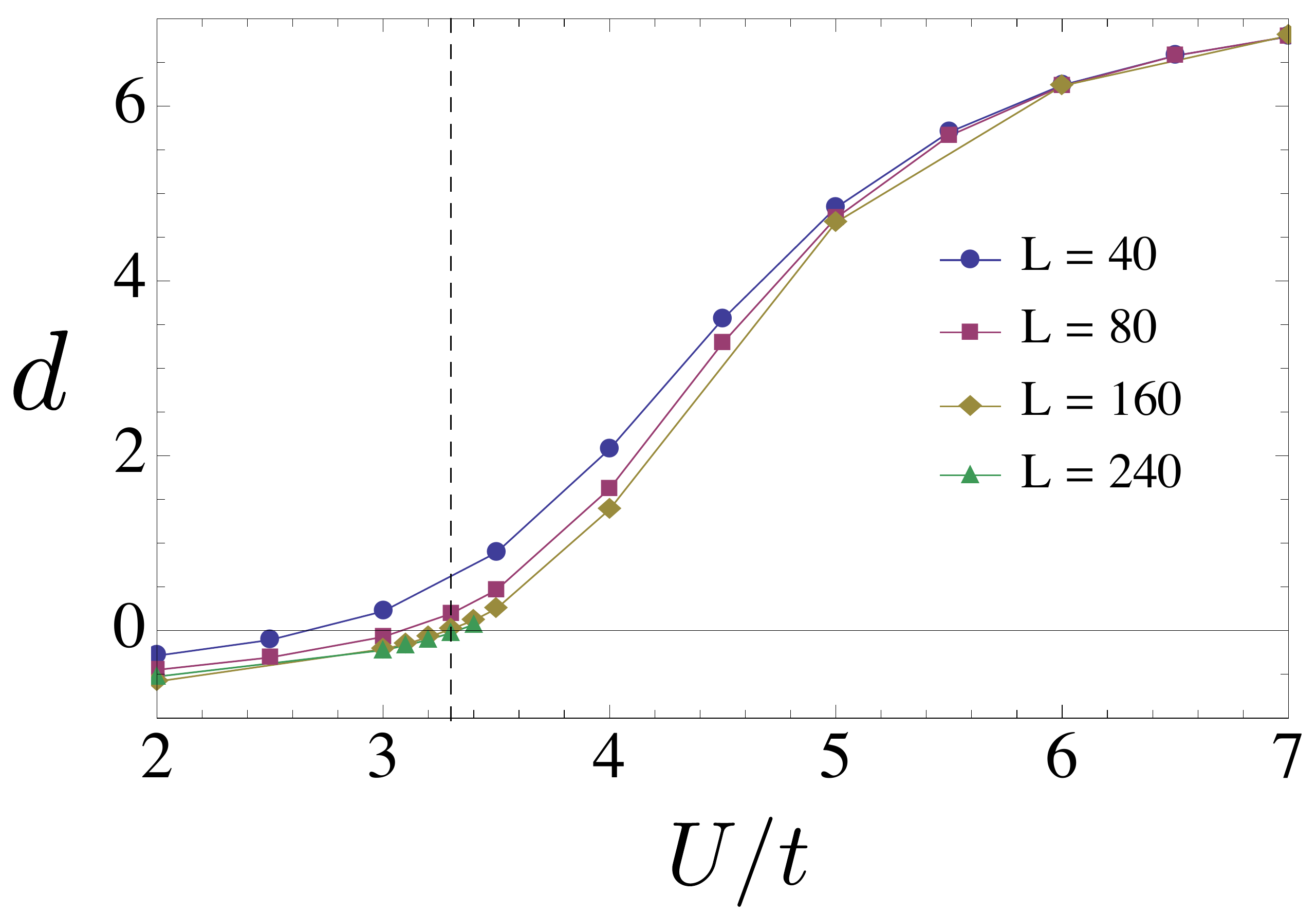}
\caption{\label{fig:BHphasetrans}
(Color online) Phase transition for the BH model: the quantity $d$, defined in the text, is plotted against $U/t$ for different lengths.  The vertical dashed line represents the critical point $U_c/t=3.3$ for the MI-SF phase transition, according to Ref.~\cite{kuhner00}.  All lattice lengths $L$ were calculated with DMRG block sizes of m=160, except for L=240 with m=200. 
}
\end{figure}

We see a deviation from a linear distribution for small $k$ close to the phase transition ($U/t\sim3.3$).  The curvature of the fitting curve increases as this deviation increases, eventually becoming a positive quadratic, for larger $U/t$, as shown for $U=5$ in Fig.~\ref{fig:U5smallk}.  The parameter that unambiguously identifies the MI-SF phase transition is simply the difference between the coefficients of the linear parts of each fit
\begin{equation}
d=a_1-a_2.
\end{equation}
This is negative in the SF critical phase and positive in the MI phase, crossing at the phase transition, as shown in Fig.~\ref{fig:BHphasetrans}.  The negative region of the function is due to the concavity of the tails of $(\Delta\hat{J})^2$ increasing deeper into the SF phase.

\begin{figure}[t]
\includegraphics[width=0.8\linewidth]{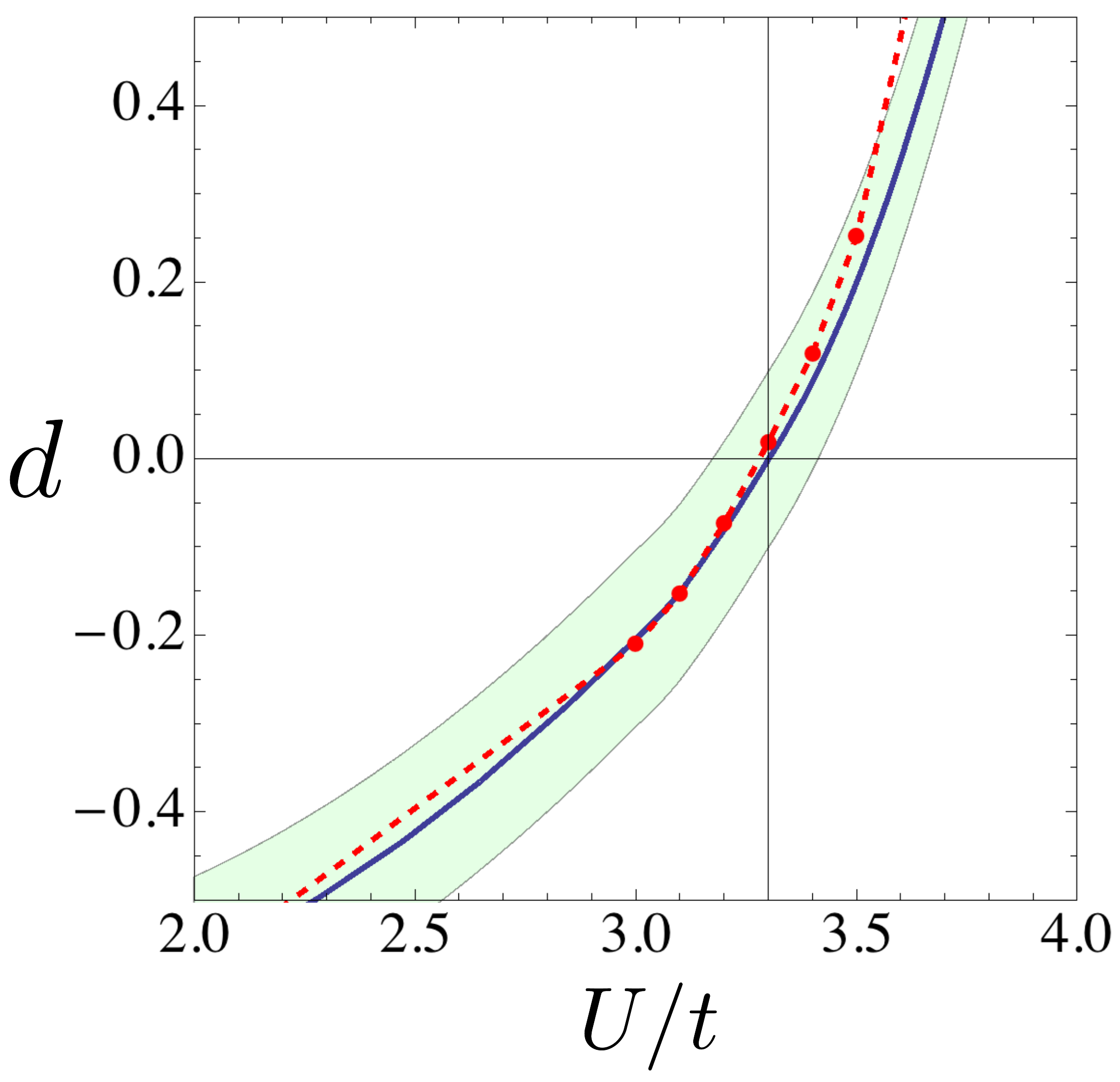}
\caption{\label{fig:threepts}
(Color online) (Blue solid) Phase transition of the BH model for $L=160$ $k=\{0.0396\pi, 0.0604\pi, 0.1\pi\}$. (Red dashed) Phase transition for $L=160$ using 100 $k$ values, for comparison and as shown in Fig.~\ref{fig:BHphasetrans}.  The shaded region indicates the error around the calculated value, determined by the dependence of the results on moving the three $k$ values. The vertical line is the estimated value $U_c/t=3.3$.
}
\end{figure}

In principle this fitting method successfully identifies the phase transition.  Although continuous tuning of the probe wavenumber $k$ is experimentally feasible \cite{Raizen} it is certainly demanding.  Therefore, we have explored the viability of using just three $k$ values to find the linear and quadratic fits and hence locate the phase transition. The three values are chosen to maximise, for large lattice sizes, the sensitivity and accuracy of the location of the critical point. The results for this more economical procedure are shown in Fig.~\ref{fig:threepts} and compared to the results obtained in Fig.~\ref{fig:BHphasetrans} for many sampling points. The shaded region indicates the error associated with this method, which takes into account the variation of the quantity $d$ when moving the three $k$ values. This variation is due to the finite size of the lattice which induces small oscillations in the signal which are suppressed when increasing the lattice size.
From the plot we estimated the location of the critical point as $U_c/t = 3.3 \pm 0.1$ which is in close agreement with the theoretical estimate $U_c/t =3.3$ \cite{kuhner00,carrasquilla13}. We notice that, as the MI-SF transition is of the Berezinskii-Kosterlitz-Thouless type \cite{fisher89}, it is quite challenging to obtain an extremely accurate location of the critical point. Nonetheless, our quantum non-demolition scheme would in practice give a fairly accurate estimate.

Although the signal for the small values of $k$ could be low in experiment, classical sources of noise could in principle be eliminated \cite{julsgaard2004}.

\section{Extended Bose-Hubbard Model: Haldane Insulator to Density Wave Transition}\label{sec:EBH}
\begin{figure}[t]
\hspace{0.5 cm}
{\bf (a)}\\
\includegraphics[width=0.85\linewidth]{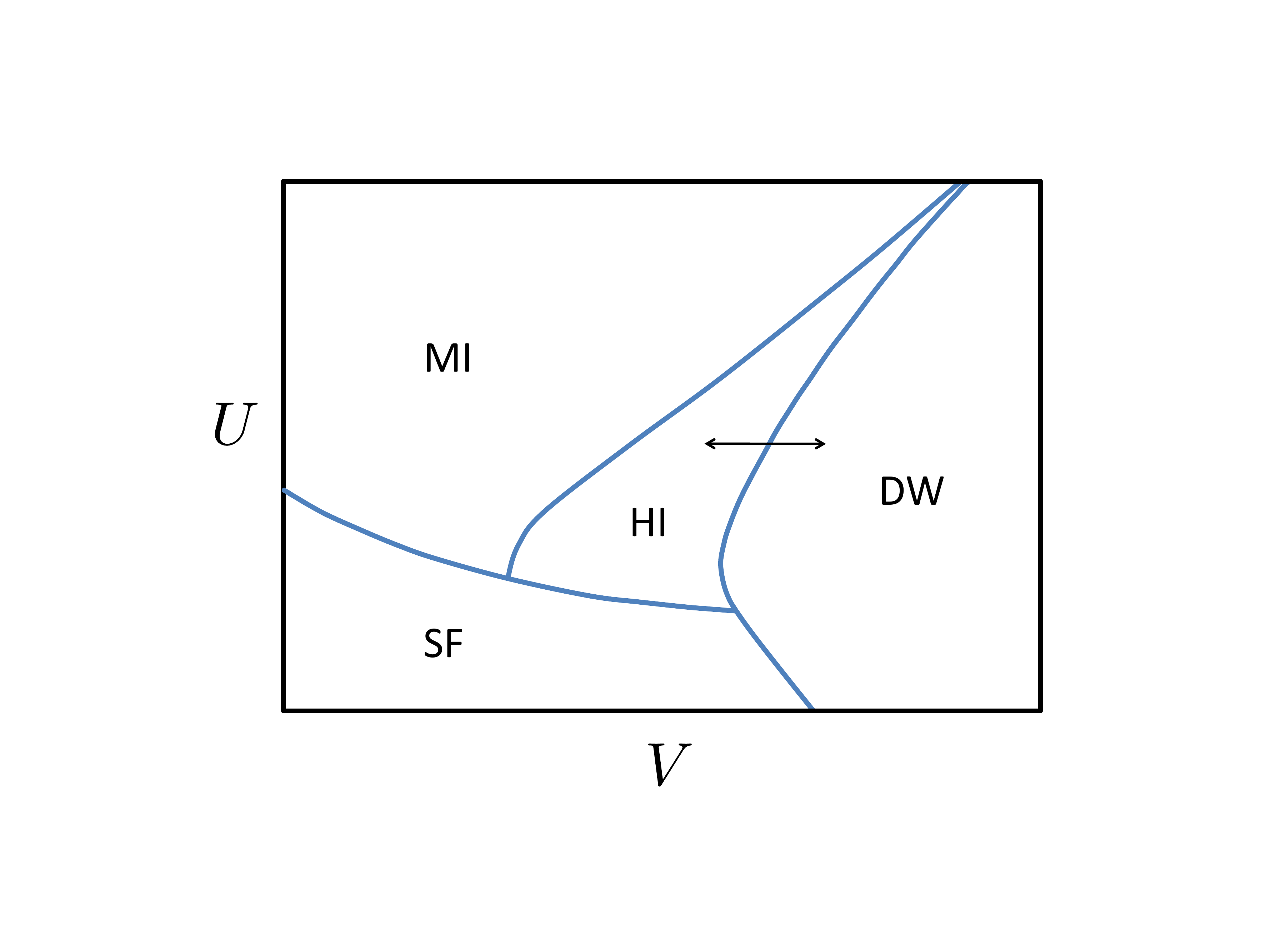}\\
{\bf (b)}\\
\includegraphics[width=\linewidth]{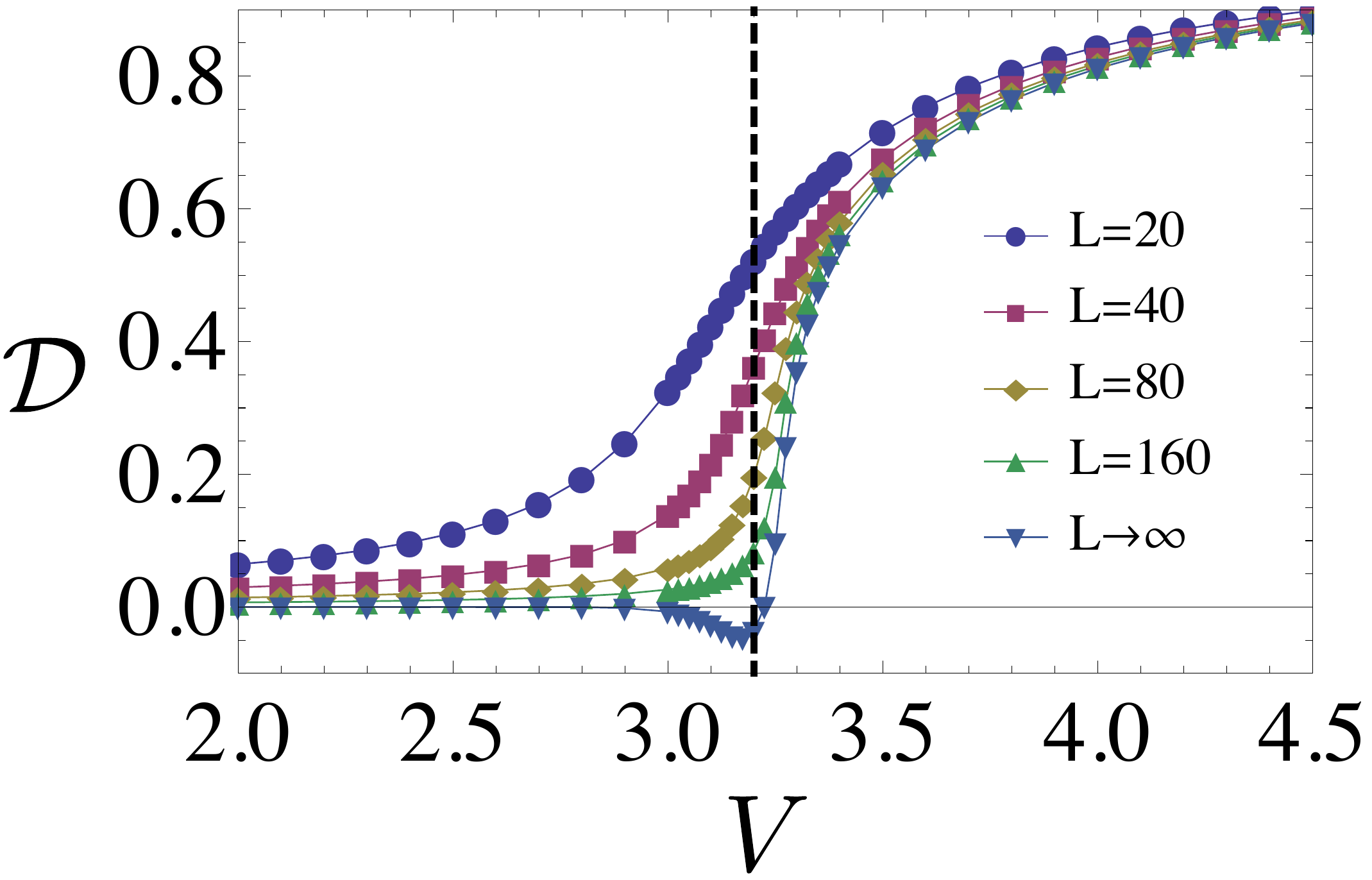}
\caption{\label{fig:EBHphasediagram}
(Color online) 
{\bf (a)} Schematic phase diagram of the extended BH model with its different phases: Mott-insulator (MI), superfluid (SF), Haldane-insulator (HI) and density wave (DW). The arrow indicates the range across the HI-DW transition that we investigate here. 
{\bf (b)} Phase transition for the EBH model.  All lattice lengths $L$ were calculated with DMRG block sizes of m=100.  The dashed line indicates the HI to DW phase transition \cite{rossini12}.}
\end{figure}

We now turn our attention to the Extended Bose-Hubbard Model, which contains an additional nearest-neighbour repulsion term with coupling $V$
\begin{equation}
\label{eq:EBHM}
\hat{H}_{EBH} = \hat{H}_{BH} + V\sum_i^L \hat{n}_i \hat{n}_{i+1}.
\end{equation}
The density-density interaction on neighbouring sites is the first approximation of a long range interaction due for example to atoms with a permanent dipole moment \cite{Pfau}.
The addition of the nearest-neighbour repulsion opens up the phase diagram to reveal two further phases: the Haldane insulator phase and the density wave phase~\cite{dallatorre,rossini12,batrouni14,kurdestany}, in addition to the MI and SF phases seen in the BHM (see Fig.~\ref{fig:EBHphasediagram}{\bf (a)}).

For unit filling and in the limit of $V\gg U$, the interaction between neighbouring sites is stronger than the on-site interaction thus favouring a double occupation on alternating sites. Thus, in the thermodynamic limit, the ground state is characterised by either double occupation on even sites and zero occupation on odd sites or vice versa. Our quantum polarization spectroscopy is especially suited to detect such periodic patterns. In the same spirit as previous publications on detection of Bose-Hubbard models using external cavities \cite{Ritsch} we consider the difference of the number of atoms at even and odd sites, $N_{even}$ and $N_{odd}$ respectively. As such, we introduce the \textit{disparity} (cf. the Appendix for details on the derivation of such a quantity):
\begin{equation}\label{eq:disparity}
\displaystyle
{\cal D} := \frac{\langle(\hat{N}_{even}-\hat{N}_{odd})^2\rangle}{L^2} = \frac{\langle\hat{J}^2\rangle}{L} - \frac{\hat{N}^2}{L^2} - \frac{2\hat{N}}{L\sqrt{L}}\langle \hat{J} \rangle , 
\end{equation}
where $\hat{N} = \hat{N}_{even} + \hat{N}_{odd}$ and we have fixed $k=\pi/2$.
In the thermodynamic limit, the disparity unambiguously identifies the quantum phase transition between the HI and DW phases, as shown in Fig.~\ref{fig:EBHphasediagram}{\bf (b)}, for $U/t=4$ as a function of $V/t$.  
The critical point, $V_c$ (denoted by the dashed line in Fig.~\ref{fig:EBHphasediagram}), is in agreement with the position of the phase transition, for $U/t=4$, found through analysis of order parameters from DMRG calculations~\cite{rossini12}.  The thermodynamic limit for a given $V$ is found through inverse-length $1/L$ extrapolation of various lattice sizes of unit filling ($N=L$). The negative disparity values shown close to the critical point are purely a result of the quadratic fit used in the inverse-length extrapolation.  Inclusion of larger lattice sizes results in the minimum flattening out as the negative values tend to zero.  

 It should be noted that the disparity cannot distinguish between the MI and the HI phases, with no features seen close to the MI-HI transition point ($V\sim 2.25 t$, for $U/t=4$).
The open question remains as to how to directly discriminate the Haldane phase. We stress that the quantity $(\Delta\hat{J})^2$ depends on two-point density-density correlations and it should not be able to detect the long range hidden order of the HI. It is possible that higher order correlations could reveal the Haldane order. We leave this study for future investigations.

\section{Superlattice Filling Transitions}\label{sec:SL}

In this section we show the versatility of collective measurements on many-body systems via QPS by applying the same measures used in the BHM case to bosons trapped in a superlattice potential.  The superlattice potential is formed by two pairs of counter-propagating laser beams of different frequencies, creating beatings and a richer structure of sublattice potentials. The 1D ground state of these systems, in the Bose-Hubbard regime, has been extensively studied \cite{roth,buonsante,Rousseau,Barthel,Hen,Dhar} and shown to give rise to fractional or non homogeneous occupations.

\begin{figure}[t]
{\bf (a)}\\
\includegraphics[width=0.9\linewidth]{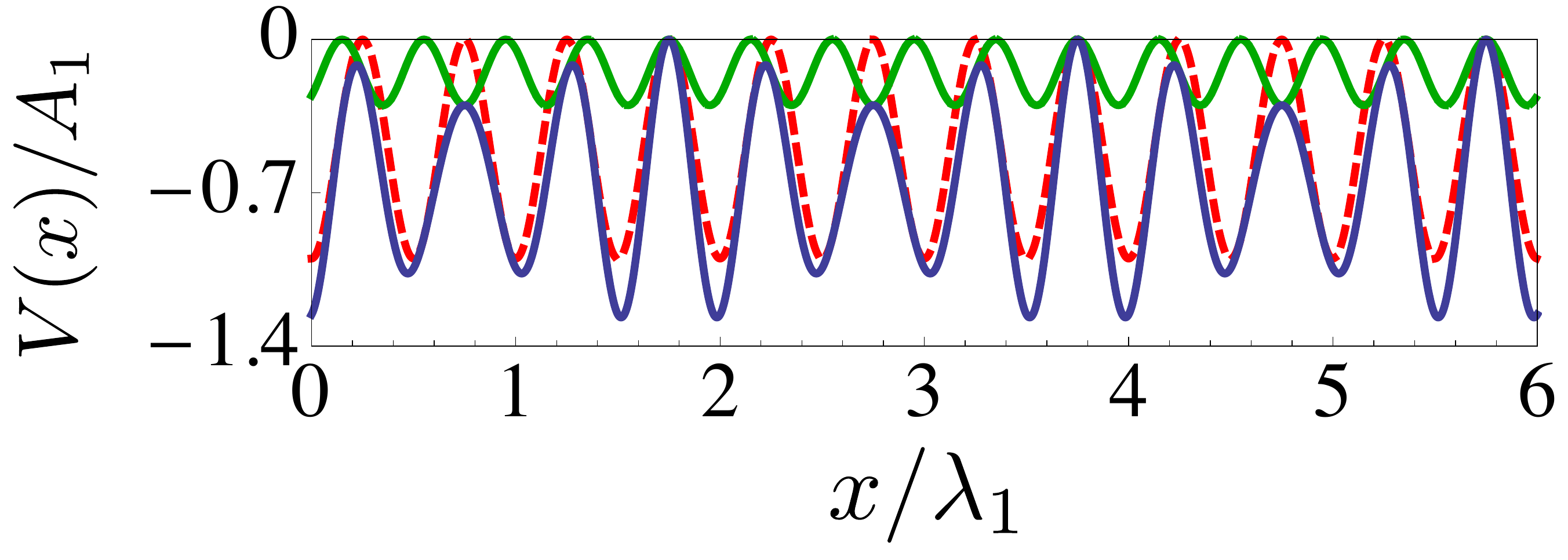}\\
{\bf (b)}\\
\includegraphics[width=0.8\linewidth]{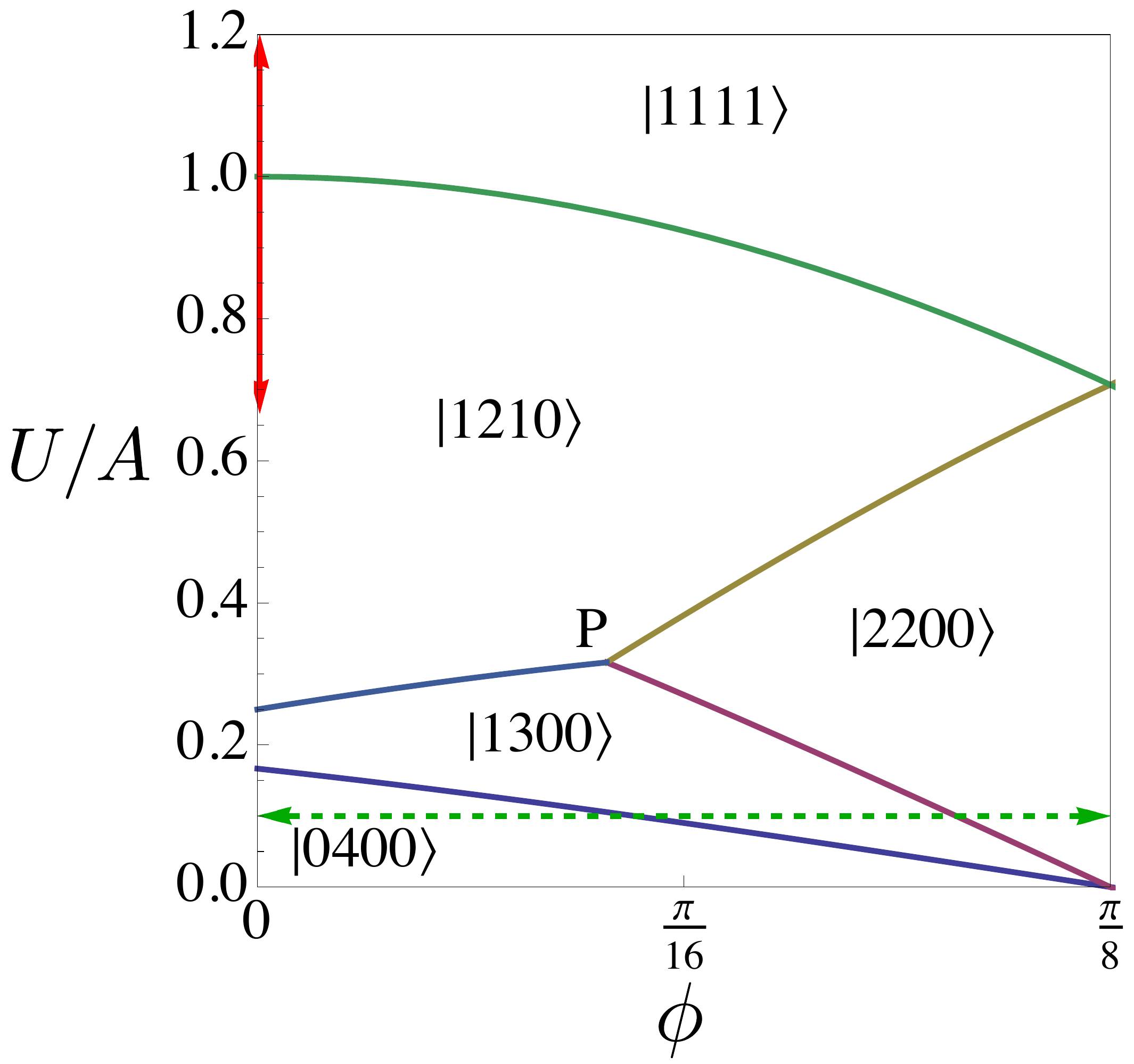}
\caption{\label{fig:SLphases}
(Color online) {\bf (a)} Superlattice potential (blue, solid) formed by two counter propagating waves, one as a larger trapping potential (red, dashed line), the other as a perturbation (green, solid line).  
{\bf (b)} Sublattice fillings for a $k_1=4/5 k_2$ superlattice potential.  The intersection point $P$ occurs at $\phi=\frac{1}{2}\arctan(\frac{1}{3})$, $U/A=\sqrt{10}/10$. The red vertical and green horizontal arrows show the two pairs of points for which we calculate the signal in Fig.~\ref{fig:SLVarJ}.}
\end{figure}

In this work we concentrate on a period-4 lattice proposed recently \cite{SLqubitgates} for implementation of atomic transistors and single and two-qubit gates.
The trapping potential has the form:
\begin{equation}
V(x) = -A_1\cos^2(k_1 x)-A_2 \cos^2(k_2 x+\phi) 
\end{equation}
where $A_1$ and $A_2$ are the relative intensities of the two lasers, $k_1$ and $k_2$ are their respective wave vectors and $\phi$ is a phase controlling the shift of the secondary lattice with respect to the first one. As in  Ref.~\cite{SLqubitgates} we choose $k_1=4/5k_2$. An example of the potential $V(x)$ is shown in Fig.~\ref{fig:SLphases}{\bf (a)}. Assuming $A_2<A_1$ we can safely assume that atoms are strongly trapped by the first laser and only perturbed by the presence of the second laser whose effect is that of shifting the local site energy. In this regime we consider the atomic Wannier functions to be localised at the minima of the potential term proportional to $A_1$. 
The second term in principle affects both the local energy and the tunnelling rate. However the effect on the latter is usually smaller and we neglect it \cite{Malte}.

Under these assumptions the Hamiltonian of the system becomes
\begin{eqnarray}
\hat{H}&=&\hat{H}_{BH}+ \sum_j \epsilon_j \hat{n}_j,
\\
\textrm{with} \quad\epsilon_j &=& -  A \cos^2\left(\frac 54 \pi j + \phi\right).
\label{eq:SLpot}
\end{eqnarray}
 Manipulation of the well depths is achieved by varying the relative amplitude $A$ that can be obtained from the corresponding overlap integral of the potential term proportional to $A_2$ and the modulus square of the atomic Wannier function. Here, in order to study a specific example, we consider this to be fixed to $A=0.3$.  Changing $\phi$ drastically changes the structure of the sublattice and hence the ground state.  
 
The ground state density distribution undergoes transitions between different configurations as the on-site repulsion $U$ and the sublattice depth $A$ compete.  Such fine control over an optical lattice clearly has great advantages, such as enabling single qubit and two qubit gates~\cite{SLqubitgates}.  
Assuming as before unit filling, i.e. one particle for each site, we first construct the zero tunnelling phase diagram. This is obtained by comparing the energies of different atomic arrangements as a function of the Hamiltonian parameters $U,A,\phi$. Owing to the periodicity of the lattice, we only need to consider 4 adjacent sites.
For example, for extremely small $U$ and $0\le \phi \le \pi/8$, all particles occupy the site with lowest energy giving a configuration $\ket{0400}$ with 4 particles in the second site. Increasing $U$, there is a critical value for which the energy for a particle to occupy the second lowest site is equal to the interaction energy of this particle to the other 3. For larger $U$ the configuration $\ket{1300}$ thus becomes favourable. The value of the on-site interaction at which this transition occurs is
\begin{equation}
U_{0400\leftrightarrow 1300} = \frac{A}{2}(\cos2\phi-\sin2\phi).
\end{equation}
We find that, under these conditions, the only possible configurations are:
\begin{equation}
\ket{0400};\quad \ket{1300};\quad \ket{1210}; \quad\ket{2200};\quad\ket{1111}
\label{eq:configs}
\end{equation}
where the last configuration corresponds to a Mott insulator, which is always obtained for sufficiently large on-site interaction energy $U$. All the transition curves between the different configurations can be obtained in a similar way.
Other possible configurations are obtained by cyclically permuting those in Eq.~\eqref{eq:configs} and appear periodically for increasing values of $\phi$. 
Thus the smallest non trivial interval to be considered is $\phi\in[0;\pi/8]$. The resulting phase diagram in the $U/A$-$\phi$ plane is shown in Fig.~\ref{fig:SLphases}{\bf (b)}.

For sufficiently large tunnelling, kinetic energy will dominate over interactions and the system will become superfluid. For sufficiently low tunnelling, we expect the superfluid to separate the different insulating phases shown in  Fig.~\ref{fig:SLphases} and disappearing only for $t\to 0$.

In this context, we show that quantum polarization spectroscopy is able to distinguish the different insulating phases. To this end we compare the signals emerging from two pairs of phases that can be obtained by changing either $U/A$ alone while keeping $\phi$ fixed or by changing $\phi$ alone while keeping $U/A$ fixed. These are:

\begin{enumerate}
\item For fixed $\phi=0$, crossing the $U/A=1$ boundary (solid red arrow in Fig.~\ref{fig:SLphases}) between the $\ket{1210}$ and $\ket{1111}$ configurations.

\item For fixed $U/A=0.12$ we consider the values $\phi=0$ and $\phi=\pi/8$, where  the ground state configurations switch from $\ket{0400}\leftrightarrow\ket{2200}$ (dashed green arrow in Fig.~\ref{fig:SLphases}).  

\end{enumerate}

\begin{figure}[t]
{\bf (a)}\\
\includegraphics[width=0.9\linewidth]{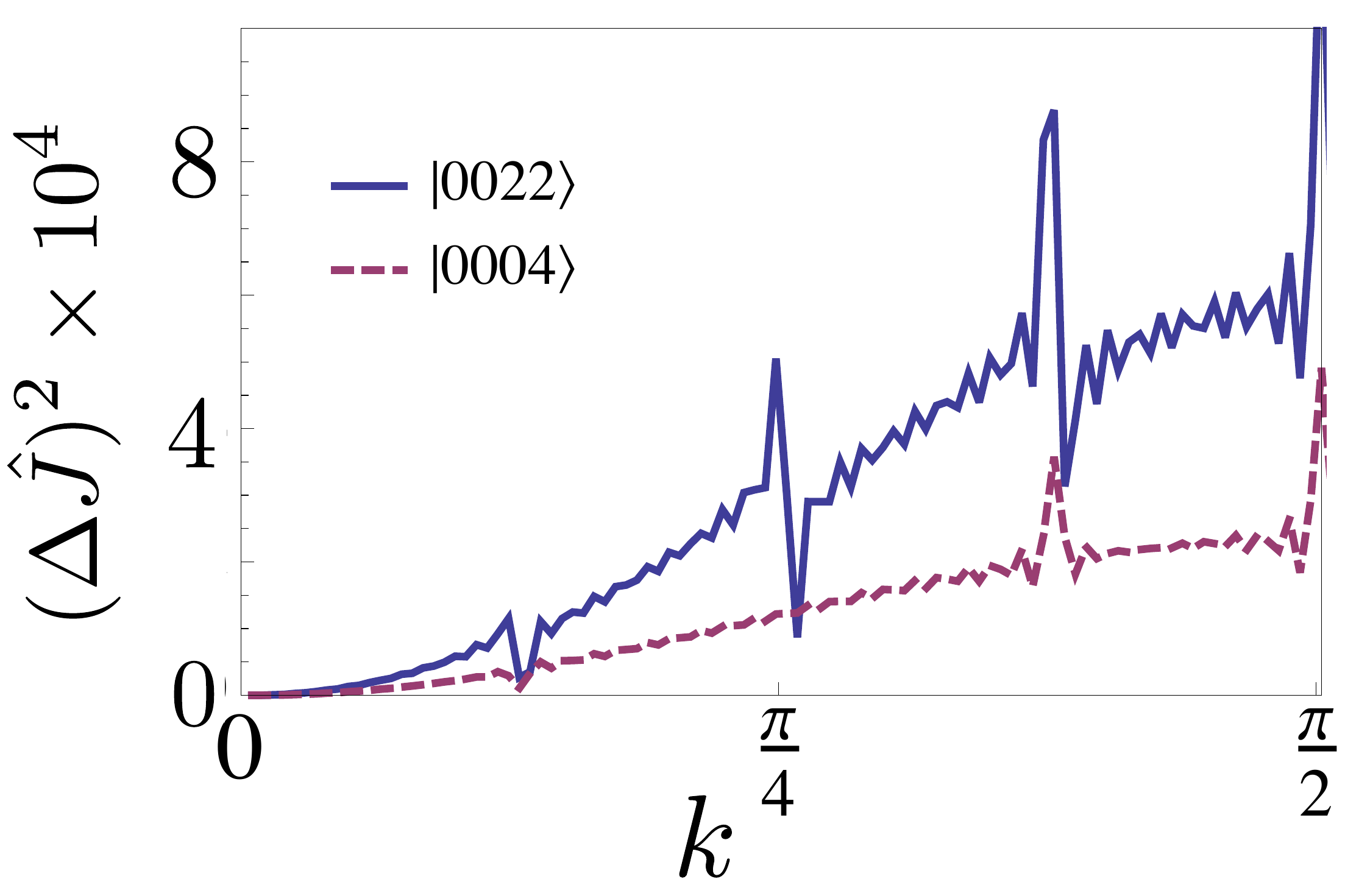}\\
{\bf (b)}\\
\includegraphics[width=0.9\linewidth]{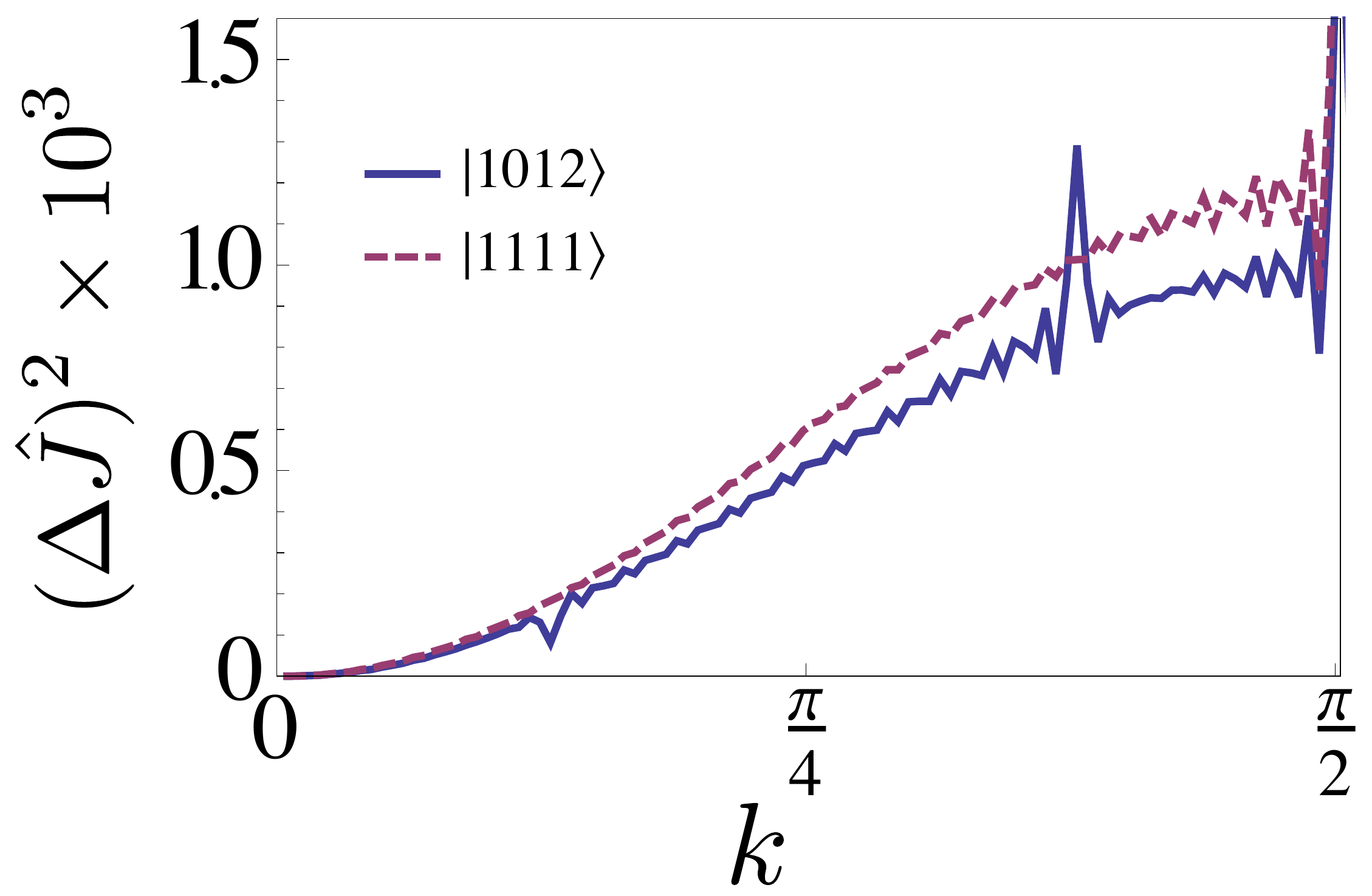}
\caption{\label{fig:SLVarJ}
(Color online) The $\hat{J}$ variance for different sublattice fillings from QPS measurements on a superlattice, for $t/U=10^{-2}$.  {\bf (a)} Variance distribution of two fillings resulting from a change in the relative phase $\Delta\phi=\pi/8$.  A clear difference can be seen between the two fillings at $k=\pi/4$.
{\bf (b)} Variance distribution of two fillings corresponding to a change in the on-site repulsion $\Delta U/A=2/3$ (scaled by the relative amplitude).  The two fillings are differentiated by the peak at $k=\pi/3$ for $\ket{1012}$.  DMRG parameters: lattice length $L=40$; block size $m=40$.
}
\end{figure}

We propose a practical method for differentiating between the different phases by measuring the variance of the effective total angular momentum, as witnessed by our DMRG results of Fig.~\ref{fig:SLVarJ} for realistic values of the tunnelling $t=10^{-2} U$. For this case the system is still insulating and in the same phases shown in Fig.~\ref{fig:SLphases}.
Figure~\ref{fig:SLVarJ}{\bf (a)} shows a clear feature for the $\ket{2200}$ state at the probe wave vector $k=\pi/4$ that does not appear for the $\ket{0400}$ state. Thus, although the two curves differ quantitatively, they also exhibit different features which make them robust for experimental implementations.  

Similarly, we analysed the transition $\ket{1012}\leftrightarrow \ket{1111}$ whose corresponding signals are shown in Fig.~\ref{fig:SLVarJ}{\bf (b)}. Once again, the $\ket{1210}$ arrangement leads to a peak at $k=\pi/3$ that is absent for the regular $\ket{1111}$ Mott state.

These examples show the power of the quantum polarization spectroscopy to detect different arrangements of atoms in superlattice potentials.

\noindent
\section{Conclusions}
\label{sec:conclusions}
We have shown how the phase transitions of the 1D Bose-Hubbard model and 1D Extended Bose-Hubbard model can be located and identified using non-demolition measurements via quantum polarization spectroscopy.  The Mott-insulator to superfluid phase transition of the 1D BHM is characterised by a comparison between a linear and quadratic fitting function, in the tail of the distribution of the variance of the collective angular momentum operator $\hat{J}$ as a function of the probe wavevector $k$.  We have obtained very accurate results compared to previous estimates of the critical point using $\sim100$ sampling points.  We have tested the robustness of the analysis in the limiting case of just three measurements and we are able to find the phase transition within the bounds found through particle and hole excitation energies, via DMRG calculations~\cite{kuhner00}.

The Haldane insulator to density wave phase transition of the EBHM is unambiguously identified by application of the {\it disparity} measure derived from the inherent difference in particle density in even and odd sites in the two phases.  Extrapolation to the thermodynamic limit shows very good agreement with calculations of the same transition found through order parameter and correlation function analysis~\cite{rossini12}. This leaves open the question on how to directly characterize the Haldane insulator using QPS, beyond just characterizing it as a phase without crystal ordering or superfluid correlations.

Finally, the versatility of the approach used to detect BHM phase transitions is tested when applied to the inhomogeneous potential arising from a superlattice.  We have shown that certain sublattice arrangements are effectively distinguishable through the QPS method since they exhibit clear signals in the variance of the total angular momentum operator. 

Our work can be further extended to more complicated Bose- or Fermi-Hubbard models in different lattice geometries, including two-dimensional ones and with other spin interactions.

\acknowledgments
We acknowledge illuminating discussions with  R. Fazio, A. L\"auchli, M. Lewenstein, D. Rossini and A. Sanpera.
This work is supported by the UK EPSRC (EP/L005026/1, EP/K029371/1, EP/G004579/1), the John Templeton Foundation (grant ID 43467), the EU Collaborative Project TherMiQ (Grant Agreement 618074).
BR thanks the Professor Caldwell Travel Studentship for support and the hospitality of the Ultracold Quantum Gases Group at Aarhus University.

~

\appendix

\section{Derivation of the Disparity}
Here we explicitly derive the disparity used in the characterisation of the EBHM.  We start by re-expressing Eq.~\eqref{eq:J} using the cosine double angle trigonometry identity:
\begin{equation}\label{Aeq:Jver2}
\hat{J} = \frac{1}{\sqrt{L}} \sum_{i=1}^L \left\{\hat{n}_i\cos\left[2k(i - \alpha)\right] + \hat{n}_i\right\}.
\end{equation}
The summation in Eq.~(\ref{Aeq:Jver2}) is split into the sum over odd sites and the sum over even sites, to give $\hat{N}_{odd}$ and $\hat{N}_{even}$, respectively, which act on different Hilbert spaces.  Additionally, if the probe wavevector is locked at $k=\pi/2$, with no displacement between the probe and the lattice ($\alpha=0$), and ensuring total particle number conservation $\sum^L_i\hat{n}_i=\hat{N}$, the collective angular momentum operator now takes the form
\begin{equation}\label{Aeq:J}
\hat{J} = \frac{\hat{N}}{\sqrt{L}} + \frac{\hat{N}_{even}-\hat{N}_{odd}}{\sqrt{L}} = \frac{2\hat{N}_{even}}{\sqrt{L}},
\end{equation}
where $\hat{N}=\hat{N}_{even}+\hat{N}_{odd}$.  By taking the mean of the square, we have
\begin{equation}
\langle\hat{J}^2\rangle = \frac{\hat{N}^2}{L} + \frac{\langle(\hat{N}_{even}-\hat{N}_{odd})^2\rangle}{L} + \frac{2\hat{N}}{\sqrt{L}}\frac{\left\langle \hat{N}_{even}-\hat{N}_{odd}\right\rangle}{\sqrt L}
\end{equation}
and by rearranging and using Eq.~(\ref{Aeq:J}),
\begin{align}
\frac{\langle(\hat{N}_{even}-\hat{N}_{odd})^2\rangle}{L} =~&\langle\hat{J}^2\rangle - \frac{\hat{N}^2}{L} - \frac{2\hat{N}}{\sqrt{L}}\left(\langle \hat{J} \rangle - \frac{\hat{N}}{\sqrt{L}}\right) \nonumber \\
=~&\langle\hat{J}^2\rangle + \frac{\hat{N}^2}{L} - \frac{2\hat{N}}{\sqrt{L}}\langle \hat{J} \rangle.
\end{align}
Since the above equation varies from $0$ to $L$ in the density wave phase, we normalise the measure to finally arrive at the definition of the disparity given in Eq.~\eqref{eq:disparity}.


\begin{thebibliography}{99}
\bibitem{review_optical_lattices}
See for example: I. Bloch, J. Dalibard, and W. Zwerger, Rev. Mod. Phys. 80, 885 (2008); M. Lewenstein, A. Sanpera and V. Ahufinger, \emph{Ultracold Atoms in Optical Lattices, Simulating Quantum Many-Body Systems}, Oxford University Press (2012); \emph{Many-Body Physics with Ultracold Gases}, Lecture Notes of the Les Houches Summer School: Volume 94, July 2010 Edited by C. Salomon, G. V. Shlyapnikov, and L. F. Cugliandolo, Oxford University Press (2012).

\bibitem{nature_insight}
AA.VV., Nature Physics Insight on \emph{Quantum Simulation}, Nature Phys. {\bf 8} (2012).

\bibitem{jaksch98}
D. Jaksch, C. Bruder, J. I. Cirac, C. W. Gardiner, and P. Zoller, Phys. Rev. Lett. {\bf 81}, 3108 (1998).

\bibitem{Greiner2002} 
M. Greiner, O. Mandel, T. Esslinger, T. W. H\"{a}nsch, and I. Bloch, Nature (London) {\bf 415}, 39 (2002).

\bibitem{fisher89}
M. P. A. Fisher, P. B. Weichman, G. Grinstein, and D. S. Fisher, Phys. Rev. B, {\bf 40}, 546 (1989).

\bibitem{dutta_review}
O. Dutta, M. Gajda, P. Hauke, M. Lewenstein, D. L{\"u}hmann, B. A. Malomed, T. Sowi{\'n}ski, and J. Zakrzewski, arXiv:1406.0181 

\bibitem{BuchlerPRL2005}
H. P. B\"uchler, M. Hermele, S. D. Huber, Matthew P. A. Fisher, and P. Zoller
, Phys. Rev. Lett. {\bf 95}, 040402 (2005).

\bibitem{Jaksch2002}
D. Jaksch, V. Venturi, J.I. Cirac, C.J. Williams, and P. Zoller, Phys. Rev. Lett. {\bf 89}, 040402 (2002).

\bibitem{Ketterle_review}
W. Ketterle, D.S. Durfee, and D.M. Stamper-Kurn:
{\it Making, probing and understanding Bose-Einstein condensates}.
In Bose-Einstein condensation in atomic gases, Proceedings of the International School of Physics ``Enrico Fermi", Course CXL, edited by M. Inguscio, S. Stringari and C.E. Wieman (IOS Press, Amsterdam, 1999) pp. 67-176.

\bibitem{atom_microscope}
J. F. Sherson, {\it et al.}, Nature {\bf 467}, 68 (2010); W. S. Bakr,  {\it et al.}, Science {\bf 329}, 547 (2010).

\bibitem{endres}
M. Endres {\it et al.}, Science {\bf 334}, 200 (2011).

\bibitem{corco}
T. A. Corcovilos, S. K. Baur, J. M. Hitchcock, E. J. Mueller, and R. G. Hulet, Phys. Rev. A {\bf 81}, 013415 (2010).

\bibitem{Ritsch}
I. B. Mekhov, C. Maschler, and H. Ritsch, Nature Phys. {\bf 3}, 319 (2007); Phys. Rev. A {\bf 76}, 053618 (2007); I. B. Mekhov, H. Ritsch,  J. Phys. B: At. Mol. Opt. Phys. {\bf 45}, 102001 (2012).

\bibitem{eckert07}
K. Eckert, L. Zawitkowski, A. Sanpera, M. Lewenstein, E.S. Polzik, Phys. Rev. Lett. {\bf 98}, 100404 (2007).
\bibitem{roscilde09}
T. Roscilde, M. Rodriguez, K. Eckert, O. Romero-Isart, M. Lewenstein, E.
Polzik, A. Sanpera, New. J. Phys. {\bf 11}, 055041 (2009).

\bibitem{dechiara11}
G. De Chiara, O. Romero-Isart, A. Sanpera, Phys. Rev. A 83, 021604(R) (2011); G. De Chiara and A. Sanpera, J. Low Temp. Phys. {\bf 165}, 292 (2011).

\bibitem{hauke13}
P. Hauke, R. J. Sewell, M. W. Mitchell, and M. Lewenstein, Phys. Rev. A {\bf 87}, 021601 (2013).

\bibitem{dallatorre}
E. G. Dalla Torre, E. Berg, and E. Altman, Phys. Rev. Lett. {\bf 97}, 260401 (2006); E. Berg, E. G. Dalla Torre, T. Giamarchi and E. Altman, Phys. Rev. B 77 245119 (2008).


\bibitem{rossini12}
D. Rossini and R. Fazio, New J. Phys. {\bf 14}, 065012 (2012).

\bibitem{SLqubitgates}
N. B. J{\o}rgensen, M. G. Bason, and J. F. Sherson, Phys. Rev. A {\bf 89}, 032306 (2014).

\bibitem{Kupriyanov}
D. V. Kupriyanov, O. S. Mishina, I. M. Sokolov, B. Julsgaard, and E. S. Polzik, Phys. Rev. A {\bf 71}, 032348 (2005).

\bibitem{HPtransformation}
T. Holstein and H. Primakoff, Phys. Rev. {\bf 58}, 1098 (1940).

\bibitem{white92}
S. R. White, Phys. Rev. Lett. {\bf 69}, 2863 (1992).

\bibitem{schollock05review}
U. Schollw{\"o}ck, Rev. Mod. Phys. {\bf 77}, 259 (2005).

\bibitem{DMRGdechiara08}
G. De Chiara, M. Rizzi, D. Rossini, S. Montangero, J. Comput. Theor. Nanosci. {\bf 5}, 1277 (2008).

\bibitem{kuhner00}
T. D. K\"uhner, S. R. White, and H. Monien, Phys. Rev. B {\bf 61}, 12474 (2000).

\bibitem{giamarchisbook}
T. Giamarchi, \emph{Quantum Physics in One Dimension}, Clarendon Press (Oxford, 2003).

\bibitem{ejima}
 S. Ejima, H. Fehske, and F. Gebhard, Europhys. Lett. {\bf 93}, 30002 (2011).

\bibitem{Raizen}
T. C. Li, H. Kelkar, D. Medellin, M. G. Raizen, 
Opt. Express {\bf 16}, 5465 (2008).

\bibitem{carrasquilla13}
J. Carrasquilla, S. R. Manmana and M. Rigol, Phys. Rev. A {\bf 87}, 043606 (2013).

\bibitem{julsgaard2004}
B. Julsgaard, et al., Nature {\bf 432}, 482 (2004); J. F. Sherson, et al., Nature {\bf 443}, 557 (2006).

\bibitem{Pfau}
A. Griesmaier, J. Werner, S. Hensler, J. Stuhler, and T. Pfau,
Phys. Rev. Lett. {\bf 94}, 160401 (2005).

\bibitem{batrouni14}
G. G. Batrouni, R.T. Scalettar, V. G. Rousseau, and B. Gr\'emaud, 	Phys. Rev. Lett. {\bf 110}, 265303 (2013).

\bibitem{kurdestany}
J. M. Kurdestany, R. V. Pai, S. Mukerjee, and R. Pandit, arxiv:1211.5202.

\bibitem{roth}
R. Roth and K. Burnett, Phys. Rev. A {\bf 68}, 023604 (2003).

\bibitem{buonsante}
P. Buonsante and A. Vezzani, Phys. Rev. A {\bf 70}, 033608 (2004);
P. Buonsante, V. Penna, A. Vezzani, 	Phys. Rev. A {\bf 70}, 061603 (2004); Laser Physics {\bf 15}, 361 (2005).

\bibitem{Rousseau}
V.G. Rousseau, D.P. Arovas, M. Rigol, F. H\'ebert, G.G. Batrouni, R.T. Scalettar, Phys. Rev. B {\bf 73}, 174516 (2006).

\bibitem{Barthel}
T. Barthel, C. Kasztelan, I. P. McCulloch, U. Schollw\"ock, 	Phys. Rev. A {\bf 79}, 053627 (2009).

\bibitem{Hen}
I. Hen, M. Rigol, Phys. Rev. B {\bf 80}, 134508 (2009).

\bibitem{Dhar}
A. Dhar, T. Mishra, R. V. Pai, and B. P. Das, Phys. Rev. A {\bf 83}, 053621 (2011).

\bibitem{Malte}
M. C. Tichy, M. K. Pedersen, K. M{\o}lmer, and J. F. Sherson,
Phys. Rev. A {\bf 87}, 063422 (2013).

\end{thebibliography}
\end{document}